\documentclass[proof,hidelinks]{WileyASNA-v1}
\articletype{Article Type}%
\received{00}
\revised{00}
\accepted{00}

\usepackage{placeins}
\usepackage{afterpage}
\newcommand{\mycomment}[1]{}

\begin{document}
\title{Apparent Stability in Self-Gravitating Turbulence and the Evolution of Molecular Clouds}
\author[1]{Eric Keto}
\authormark{Eric Keto}
\titlemark{Molecular Clouds}
\address[1]{\orgdiv{Department of Astronomy}, \orgname{Harvard University},
\orgaddress{\state{Cambridge, MA}, \country{USA}}}
\corres{Eric Keto}
 \presentaddress{Center for Astrophysics, 60 Garden St, Cambridge, MA 02138}
 
 
 \abstract[Abstract]{
Recent observations of hydrostatic structure and virial equilibrium in supersonically turbulent, self-gravitating molecular clouds imply a stability that contrasts with the transcience of turbulent structure. To investigate this contradiction, we model a molecular cloud as a turbulent eddy and study its evolution as a dynamical system. In a two-dimensional phase space of structure and energy, we find that the equilibrium is a saddle point, stable in the direction aligned with force balance, but unstable in the direction of energy balance because of the combination of the turbulent dissipation and the negative heat capacity of self-gravitation. Near the saddle point, evolutionary trajectories follow a characteristic pattern that first approaches the equilibrium before departing in the direction of instability. Since the phase-space speed is proportional to the virial and energy imbalance, trajectories slow near the equilibrium resulting in a local overdensity of clouds. Also, near equilibrium, the relaxation to force balance is faster than the growth rate of the instability in energy. Consequently, more clouds are observed in near equilibrium states with hydrostatic structure even though the equilibrium is metastable. This resolves the apparent contradiction of equilibrium structure observed in dynamically unstable, self-gravitating turbulence.
}
 
 
 \maketitle

  \section{Introduction}
  
 Molecular clouds are transient density fluctuations in the supersonic, self-gravitating turbulence of the molecular interstellar medium (ISM). However, recent observational analyses of clouds in the Galactic Ring and in the galaxy M31 show that their internal density structure is well described by solutions of the isothermal Lane--Emden equation for hydrostatic equilibrium \citep{Keto_2024,Keto_2025,Lada_2025}. These results based on new differential methods of analysis rather than the integral method of virial analysis are especially compelling because they do not depend on the absolute calibration of the cloud mass through a CO-to-H$_2$ conversion factor, nor do they require a physically unique definition of the cloud boundary. The observational result of internal force balance therefore appears robust, and a substantial fraction of molecular clouds are structured in virial and hydrostatic equilibrium. 
These observations sharpen a basic theoretical question. 
How does equilibrium structure develop within the fluctuations of supersonic, self-gravitating turbulence?

A recent thermodynamic treatment describes self-gravitating turbulent clouds in terms of a Gibbs potential in a two-dimensional phase space of density and energy \citep{Donkov_2025b}. The manifold defined by force and energy balance has a saddle structure at equilibrium. The stable direction is associated with force balance while the unstable direction is associated with the energy balance. This alignment of the saddle 
raises the question 
whether the observations of equilibrium structure may be
revealing the stability of the force balance  while the overall dynamical equilibrium is
driven by an instability in energy.
This question contrasts two existing interpretations of the dynamics of the molecular interstellar medium (ISM). 

In the hypothesis of gravoturbulent
fragmentation \citep{Klessen_2000,Padoan_2002,MacLow_2004}, compressed regions either gravitationally collapse or disperse depending on whether the compressed
density exceeds a Jeans-type threshold. The fate of a cloud is dependent on its 
boundary conditions -- the external compression -- and decided by an
instability that operates along one direction in phase space specified as density. 
In this view there is no evolutionary path toward an equilibrium density structure. 

However, the thermodynamics suggests that cloud evolution cannot be understood from force balance alone. 
Rather, the stability of a cloud depends on the interplay between structural support and the evolution of its turbulent energy reservoir.

A complementary interpretation is provided by a dynamical
systems analysis of cloud evolution proposed in \citet{Keto_2020} (KFB).
Reformulated as a driven system with boundary conditions consistent with a turbulent cascade, 
we find that the equilibrium has a spiral saddle structure, stable in force
balance and unstable in energy balance, in agreement with the thermodynamic description.

The local geometry of the phase space around equilibrium has an
immediate observational consequence. A broad class of trajectories
near the saddle first moves toward the equilibrium as the restoring
dynamics reduce the structural or force imbalance. The trajectories
then turn away when the slower energetic instability becomes
dominant.

Because the phase-space speed depends on the combined magnitudes of the virial 
and energy imbalances, trajectories slow as they approach equilibrium. Conservation of phase-space flux therefore creates an overdensity of trajectories in the neighborhood of the equilibrium. In a statistical ensemble, this overdensity implies that a larger fraction of clouds will be found in approximate hydrostatic and virial equilibrium, even though the equilibrium itself is not fully stable. In this way, the prevalence of apparently equilibrated clouds can be reconciled with cloud lifetimes that remain comparable to the timescale of the energy instability which is the turbulent dissipation timescale. 

The two interpretations, gravoturbulent fragmentation and the metastable equilbrium of the thermodynamics and system dynamics,  
present a straightforward observational discrimination. Hydrostatic and virial structure are precluded
in the one-dimensional interpretation of gravoturbulent fragmentation  but predicted by the structure of the two-dimensional metastable equilibrium.  
We further this conclusion in the remainder of this present study.

This dynamical interpretation also clarifies the relationship among previous descriptions of self-gravitating turbulence. The analysis of \citet{Elmegreen_1993}, which informs the hypothesis of gravoturbulent fragmentation, operates on a lower-dimensional slice through the full two-dimensional cloud phase space. Likewise, the observationally motivated, modified Larson relations of \citet{Heyer_2009} reside on a projection of the three-dimensional phase space of virial energies onto a two-dimensional plane of observables.

\section{Thermodynamic and Dynamical Descriptions of Near Equilibrium Self-Gravitating Clouds}\label{Background}

\subsection{Thermodynamic description of a two-dimensional equilibrium manifold}

\citet{Donkov_2025b} formulate a thermodynamic description of turbulent, self-gravitating molecular clouds by introducing a macrostate defined by a temperature-like quantity, $\theta$, associated with turbulent kinetic energy and a density, $n$, describing the spatial structure of the cloud. Within this framework, the first law of thermodynamics is applied to a fluid element, leading to expressions for the entropy and thermodynamic potentials as functions of the macroscopic variables. In particular, for a cloud embedded in an environment characterized by fixed temperature and pressure, the relevant potential is the Gibbs free energy (Donkov et al. 2025b, \S 3.4).

The off-equilibrium Gibbs potential $g_0$ is
\begin{equation}
g_0(\theta,n) = \varepsilon(\theta,n) - \theta_0 s(\theta,n) + P_0 \left(\frac{1}{n}\right),
\end{equation}
where $\epsilon,\ s,\ P$ are the total energy, entropy, and pressure. The second derivatives evaluated at equilibrium
\begin{equation}
\frac{\partial^2 g_0}{\partial \theta^2} < 0,
\qquad
\frac{\partial^2 g_0}{\partial n^2} > 0,
\end{equation}
indicate a saddle point \citep[][S 3.43]{Donkov_2025b}. Thus, the system is stable with respect to density (structural) perturbations but unstable with respect to temperature (turbulent energy) perturbations.

We may equivalently describe the cloud by its radius $R$ and turbulent kinetic energy proportional to the 
one-dimensional turbulent velocity dispersion squared $K \propto \sigma^2$ . The saddle structure found by 
\citet{Donkov_2025b} in $(n,\theta)$ space is therefore equivalent to a saddle in the observable $(R,K)$ space with positive curvature in the structural direction and negative curvature in the energetic direction.
This result implies that molecular clouds can achieve approximate equilibrium configurations, but these configurations are intrinsically metastable rather than true minima of the thermodynamic potential.

\subsection{Dynamical systems interpretation}

 \citet{Keto_2020} (KFB) describe cloud evolution as a dynamical system in the $(R,K)$ phase space by combining  the first law of thermodynamics, the time-dependent virial equation, and the energy loss rate in turbulent dissipation (KFB \S 2.2),
\begin{equation}\label{first_law}
\frac { dQ } { dt } = 2K \frac{ d } { dt } \log{\sigma R},
\end{equation}
\begin{equation}
\frac { d^2(R^2) } { dt^2 } = \frac{ 6 } { \beta }  \left( \sigma^2 - \frac { \Gamma GM^2 } { 3 } \right) ,
\end{equation}
and
\begin{equation}
\frac { dK } { dt } = -\frac { K } { t_D }
\end{equation}

Here $\beta $ is a geometric factor related to the moment of inertia with a value of 0.3 for an $R^{-2}$
density structure (Lane-Emden) and 0.6 for uniform density, and $\Gamma$ is also a geometric
factor with a value of 3/5 for uniform density and 0.73 for a Lane-Emden structure; $G$ and $M$
refer to the gravitational constant and cloud mass.

The quantity $\log (\sigma R)$ in equation \ref{first_law} is interpreted as a turbulent entropy
with $W \equiv \sigma R$ analogous to the number of microstates.
The same entropy is found in the thermodynamic description of \citet{Donkov_2025b}. 

Expressed as a system of coupled ordinary equations in dimensionless form (KFB eqs.\ 28, 29, 21):
\begin{equation}\label{KFB_28}
\frac{dr}{d\tau} = v,
\end{equation}
\begin{equation}\label{KFB_29}
\frac{dv}{d\tau}
=
\frac{1}{2r}
\left[
-2v^2 + \frac{6}{\beta}
\left(
\frac{w^2}{r^2} - \frac{1}{r}
\right)
\right],
\end{equation}
\begin{equation}\label{KFB_21}
\frac{dw}{d\tau} = -\frac { 1 } { 2\gamma } \frac{w^2}{r^2},
\end{equation}\\
where $s,r,w,\tau$ are the non-dimensional forms of $\sigma, R, W, t$  normalized by
their values at equilibrium and the equilibrium crossing time as appropiate, and 
$\gamma = t_D / t_X$ is the ratio of the turbulent dissipation timescale to the crossing time.

Linearizing these equations about equilibrium, KFB derive a cubic dispersion relation for the growth rate,
with one real root corresponding to monotonic growth or decay,
and a complex conjugate pair corresponding to oscillatory modes.

KFB identify the oscillatory modes with radial oscillations of the cloud on the gravitational timescale with frequency,
\begin{equation}
\varpi \sim \sqrt{\frac{GM}{R^3}} \sim \frac { 1 } { t_\mathrm{dyn} },
\end{equation}
for mass $M$, while the real root corresponds to the evolution of the turbulent energy through dissipation,
\begin{equation}
\lambda \sim \frac { 1 } { t_D }.
\end{equation} 
This describes stable structural oscillations about a state
of virial balance that drifts with the persistent dissipation of turbulent energy. 
This anticipates the saddle-point interpretation from the thermodynamic description.

However, the system does not admit a two-dimensional equilibrium
with both $dv/d\tau=0$ and $dw/d\tau=0$ in equations \ref{KFB_29} and \ref{KFB_21}
explained as follows.
The net energy loss rate is specified by the parameter $\gamma$ which allows the
input of mechanical energy to modify the loss rate.
Exact balance between
injection and dissipation corresponds to the limiting case
$1/\gamma\rightarrow0$, or $\gamma\rightarrow\infty$. In this limit
the energy equation becomes degenerate. For any finite
$\gamma$ the turbulent entropy or energy continuously decreases.
However, this is a result imposed by a mathematical limitation
of the model rather than a requirement of the physics.

\subsection{Inclusion of turbulent energy injection}\label{Pressure}

To allow for an equilibrium state, we introduce a separate source term representing energy injection from the turbulent cascade. 
The modified evolution equation for $w$ becomes
\begin{equation}
\frac{dw}{d\tau} =  -\frac { 1 } { 2\gamma } \frac{w^2}{r^2} + I(w,r),
\end{equation}
where $I$ is the rate of turbulent energy injection.

We parameterize the injection in terms of the turbulent cascade, assuming that velocity fluctuations follow a scaling relation with length $\ell$,
\begin{equation}
s(\ell) \propto \ell^\alpha,
\end{equation}
with
\begin{equation}
\alpha =
\begin{cases}
1/3 & \text{(Kolmogorov cascade)}, \\
1/2 & \text{(Larson-type scaling)}.
\end{cases}
\end{equation}
At the scale of the cloud, this implies
\begin{equation}
w = s r \propto r^{1+\alpha}.
\end{equation}
The turbulent cascade transfers energy from larger scales into the cloud at a rate proportional to the turnover time of the larger eddies, so that
\begin{equation}
I \sim \frac{w}{\tau_{\rm inj}},
\qquad
t_{\rm inj} \sim \frac{r}{s} \propto r^{1-\alpha}.
\end{equation}
Thus the injection rate may be written as
\begin{equation}
I(w,r) = I_0\,\frac{w}{r^{1-\alpha}},
\end{equation}
where $I_0$ is a dimensionless normalization that sets the strength of energy input relative to dissipation.
The evolution equation for $w$ (equation \ref{KFB_21}) therefore becomes
\begin{equation}\label{nd_energy_dissipation}
\frac{dw}{d\tau}
= -\frac { 1 } { 2\gamma }  \frac{w^2}{r^2}
+
I_0\,\frac{w}{r^{1-\alpha}}.
\end{equation}
In this case $\gamma$ specifies only the turbulent dissipation time with respect to the crossing time
rather than the net energy loss rate which is now given by the sum of both terms on the right-hand-side
of equation \ref{nd_energy_dissipation}

\subsection{Inclusion of external pressure}

In the original KFB formulation, the virial balance includes only kinetic and gravitational terms. However, observations of Galactic and extragalactic molecular clouds indicate that clouds are confined by an external pressure set by the surrounding turbulent interstellar medium \citep{Keto_2024, Keto_2025}. In these studies, the boundary of a cloud is identified with an arbitrary point on a Lane--Emden profile, and the boundary pressure is
\begin{equation}
P_b = \rho_b \sigma_b^2,
\end{equation}
for density $\rho_b$ and turbulent velocity dispersion $\sigma_b$ at the boundary.

For a homologously evolving cloud, the ratio of boundary pressure to mean internal pressure remains approximately constant. We therefore write
\begin{equation}
P_0 = \eta\,\bar{\rho}\,\sigma^2,
\end{equation}
where $\eta$ is a constant of order unity determined by the choice of boundary on the Lane--Emden solution.

The external pressure term in the virial theorem contributes
\begin{equation}
3P_0 V = 3\eta M \sigma^2,
\end{equation}
which has the same scaling as the internal kinetic energy. Thus, the effect of external pressure can be incorporated by modifying the effective kinetic term:
\begin{equation}\label{K_eff}
K_{\rm eff} = (1+\eta) K.
\end{equation}

The equation of motion for $v$ (equation \ref{KFB_29}) therefore becomes
\begin{equation}\label{nd_force_balance}
\frac{dv}{d\tau}
=
\frac{1}{2r}
\left[
-2v^2
+
\frac{6}{\beta}
\left(
(1+\eta)\frac{w^2}{r^2}
-
\frac{1}{r}
\right)
\right].
\end{equation}

\subsection{A truly constant external pressure}

In most observations of molecular clouds, the boundary of a cloud is defined by an arbitrary
threshold column density that sensibly encompasses a region of high column density 
identifiable as a cloud. The recent observations analyzed by differential methods indicate that
molecular emission generally extends beyond a threshold-selected boundary and that its column density continues to follow
the Lane-Emden structure of the cloud within the boundary. This reflects the physical
reality that molecular clouds are regions of high density within a continuous turbulent field.
The definition of external pressure adopted in the section above is consistent with this description.

Without reference to a specific model, the pressure term in the virial theorem is often written as $3P_\mathrm{ext}R^3$ which is
interpreted as a constant external pressure that acts through an $R^3$ dependence. This is
the model of a Bonnor-Ebert system and could correspond to a molecular cloud confined by a low-density,
hot ionized gas with a sound speed high enough to maintain a near constant external pressure.

Both definitions of external pressure result in an equilibrium with a saddle structure in the two dimensional 
phase space, and in both cases the trajectories of evolution first approach the point of equilibrium 
before evolving in the unstable direction. 
However, a pressure term with an $R^3$ dependence modifies the trajectories from spiral (\S\ref{trajectories})
to hyperbolic with stable and unstable branches. Calculations with this model result in the same 
conclusions as with the Lane-Emden boundary pressure.
Since this model does not apply to most molecular
clouds, for conciseness it is not discussed here.

\section{The Jacobian and eigenvalues of the dynamical system at equilibrium}

The first step in our analysis is to determine the local stability of the equilibrium point. This is achieved by constructing the Jacobian matrix of the dynamical system and evaluating it at equilibrium. The eigenvalues of the Jacobian determine the nature of the fixed point, whether perturbations grow, decay, or oscillate. In particular, the characteristic equation derived from the Jacobian yields a cubic polynomial whose roots correspond to the growth rates of perturbations. 

The corresponding eigenvectors  identify the stable and unstable directions of the equilibrium manifold and thus determine how trajectories approach and depart from equilibrium. In the present context, they distinguish between structural (radius-dominated) and energetic (turbulence-dominated) modes of evolution.
\subsection{The Jacobian}

The equilibrium point $(r_*, v_*, w_*)$ satisfies
\begin{equation}
v_* = 0,
\label{E5}
\end{equation}

\begin{equation}
(1+\eta)\frac{w_*^2}{r_*^2} = \frac{1}{r_*},
\label{E6}
\end{equation}

\begin{equation}
\frac{1}{2\gamma}\frac{w_*^2}{r_*^2}
=
I_0\,\frac{w_*}{r_*^{\,1-\alpha}}.
\label{E7}
\end{equation}

Equation \ref{E6} is the condition of virial force balance, modified by the 
external pressure term. Equation \ref{E7} is the condition that turbulent dissipation is balanced by turbulent energy injection.

Evaluating the Jacobian at equilibrium gives
\begin{equation}
J_* =
\begin{pmatrix}
0 & 1 & 0 \\[6pt]
a & 0 & b \\[8pt]
c & 0 & d
\end{pmatrix},
\label{E10}
\end{equation}
with
\begin{equation}
a = -\frac{3}{\beta r_*^3},
\label{E11}
\end{equation}

\begin{equation}
b = \frac{6(1+\eta)w_*}{\beta r_*^3},
\label{E12}
\end{equation}

\begin{equation}
c = \frac{1+\alpha}{2\gamma}\frac{w_*^2}{r_*^3},
\label{E13}
\end{equation}

\begin{equation}
d = -\frac{w_*}{2\gamma r_*^2}.
\label{E14}
\end{equation}

Since 
$a < 0, \ b > 0, \  c > 0, \ {\rm and} \ d < 0$,
the equilibrium is mechanically restoring in the $r$-direction but unstable in the energetic direction, 
as expected from the saddle-point interpretation of the Gibbs potential.

\subsection{Eigenvalues}

The three eigenvalues are the roots of the characteristic polynomial,
\begin{equation}
\lambda^3 + A\lambda^2 + B\lambda + C = 0,
\label{E20}
\end{equation}
with
\begin{equation}
A = \frac{w_*}{2\gamma r_*^2}, 
\qquad 
B = \frac{3}{\beta r_*^3}, 
\qquad 
C = -\frac{3(1+2\alpha)w_*}{2\beta\gamma r_*^5}.
\label{E21}
\end{equation}
We define the depressed cubic
\begin{equation}
\lambda = y - \frac{A}{3},
\label{E22}
\end{equation}
so that
\begin{equation}
y^3 + py + q = 0,
\label{E23}
\end{equation}
where
\begin{equation}
p = B - \frac{A^2}{3},
\label{E24}
\end{equation}
\begin{equation}
q = \frac{2A^3}{27} - \frac{AB}{3} + C.
\label{E25}
\end{equation}
The discriminant is
\begin{equation}
\Delta = \left(\frac{q}{2}\right)^2 + \left(\frac{p}{3}\right)^3.
\label{E26}
\end{equation}

For $\Delta > 0$, the roots consist of one real eigenvalue and one complex conjugate pair:
\begin{equation}
\lambda_1 =
-\frac{A}{3}
+
\left(-\frac{q}{2}+\sqrt{\Delta}\right)^{1/3}
+
\left(-\frac{q}{2}-\sqrt{\Delta}\right)^{1/3},
\label{E27}
\end{equation}

\( \displaystyle
\lambda_{2,3} =
-\frac{A}{3}
-
\frac{1}{2}
\left[
\left(-\frac{q}{2}+\sqrt{\Delta}\right)^{1/3}
+
\left(-\frac{q}{2}-\sqrt{\Delta}\right)^{1/3}
\right]
\)

\begin{equation}
\pm
\frac{i\sqrt{3}}{2}
\left[
\left(-\frac{q}{2}+\sqrt{\Delta}\right)^{1/3}
-
\left(-\frac{q}{2}-\sqrt{\Delta}\right)^{1/3}
\right].
\label{E28}
\end{equation}

The real root describes monotonic evolution in the energetic direction, while the complex conjugate pair describes oscillatory motion in the structural direction.

\subsection{Eigenvectors and the saddle structure}

The eigenvectors can be written as
\begin{equation}
\mathbf{e}_j \propto
\begin{pmatrix}
1 \\
\lambda_j \\
\dfrac{c}{\lambda_j - d}
\end{pmatrix}.
\label{E32}
\end{equation}

The eigenvectors define the geometry of the flow in phase space near the equilibrium. The positive real eigenvector defines the unstable direction along which trajectories depart from equilibrium. This corresponds to the KFB instability in entropy. 
The complex conjugate pair defines the two-dimensional stable spiral plane in which trajectories approach equilibrium through damped oscillations 
that correspond physically to gravitational breathing modes of the cloud. 
These oscillations reflect the restoring force between pressure and self-gravity.

\begin{figure*}[hbt!]
\includegraphics[width=6.5in,trim={0 0.0in 0.0in 0.0in},clip] {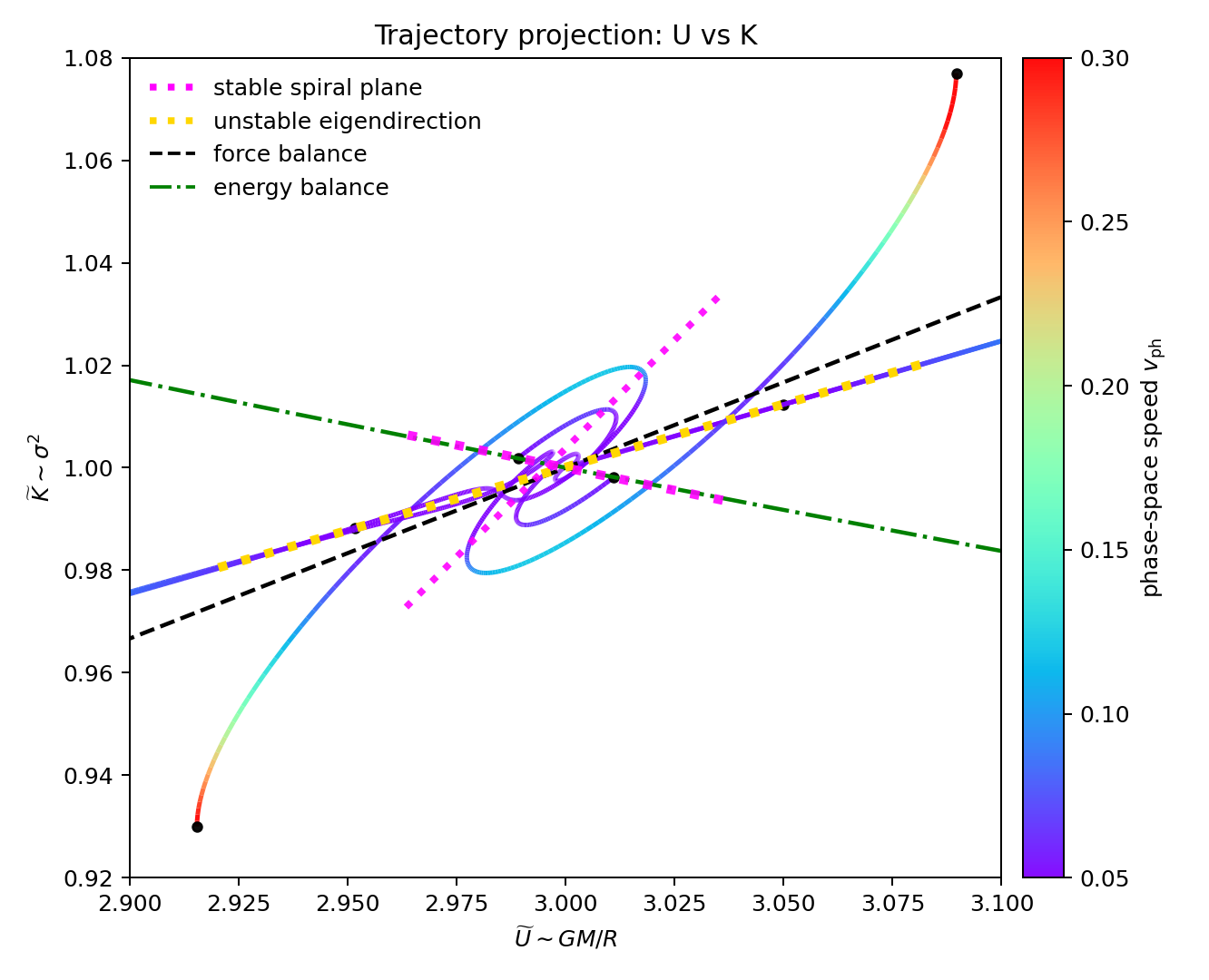}
\caption{Trajectories in non-dimensional phase space as described in \S \ref{trajectories}. The trajectories 
are projected onto the plane of the non-dimensional kinetic versus gravitational energies.}
\label{GE_KE}
\end{figure*}

\section{Numerical Solutions and Phase-Space Trajectories}\label{trajectories}

While linear stability analysis characterizes the behavior of perturbations near equilibrium, the global evolution of clouds requires
solution of the full dynamical system. We therefore compute representative trajectories in phase space, which describe the time-dependent evolution of clouds from a range of initial conditions. These trajectories illustrate the characteristic behavior expected near a saddle point: a rapid approach toward equilibrium along the stable direction, followed by slower departure along the unstable direction.

Figure \ref{GE_KE} presents the numerical solutions of the dynamical system projected onto the plane of 
non-dimensional gravitational and kinetic energies, $(\widetilde{U}, \widetilde{K})$, where
\begin{equation}
\tilde{U} \propto U \sim \frac{GM}{R}, \qquad \tilde{K} \propto K \sim \sigma^2.
\end{equation}
This projection provides a direct visualization of the virial balance between self-gravity and turbulent support.
Several trajectories are shown in figure \ref{GE_KE}, each initialized at a different point (black dots) in the $(\tilde{U},\tilde{K})$ plane. 

The force balance curve (black dashed line),
$
2\tilde{K}_{\rm eff} + \tilde{U} = 0,
$
defines virial or hydrostatic equilibrium.
The energy balance curve (green dashed line), 
$
\dot{ \tilde{K} } = 0,
$
defines energy equilibrium
where turbulent energy injection balances dissipation.
Their intersection 
$
(\tilde{U}, \tilde{K}) = (3,1)
$
corresponds to the saddle point in the two-dimensional phase space.

The projection of the stable spiral plane given by the pair of complex conjugate eigenvalues 
is shown by the two dotted magenta lines that cross
at the equilibrium. The spiral pattern followed by the trajectories approaching equilibrium aligns with 
this plane. The initial values are chosen to lie on this plane so that the trajectories
present a simple spiral in projection for illustration. From initial values off the plane, the 
trajectories would also spiral toward
the equilibrium but show as complex looping patterns in projection.

The direction of the unstable eigenvector is shown by the dashed yellow line. The trajectories
leave the equilibrium region in this direction. 

Near a saddle-point equilibrium, trajectories typically first approach the equilibrium neighborhood along the
directions of the stable modes and then depart from the neighborhood in the unstable direction before
actually reaching the equilibrium. The
exact paths depend on the system. In the modified KFB system, the virial imbalance drives a coupled 
change in $R$ and $K$ that reduces the force residual and brings the cloud closer to equilibrium. 
Once the force imbalance becomes small compared to the energy imbalance, 
the evolution of a trajectory becomes dominated by the energy imbalance. Because the
equilibrium is unstable with respect to an imbalance in energy, the system drifts 
away from equilibrium in the direction of the unstable mode.

\section{Residence time near equilibrium}

A central question is the duration for which a cloud remains near equilibrium as this determines the likelihood of observing clouds in approximate virial or hydrostatic balance. We estimate this residence time using three complementary approaches. \\
(1) From the dynamical timescale along trajectories. 
We evaluate the speed of motion along individual trajectories in phase space to determine how long a cloud remains in the vicinity of the equilibrium point. \\
(2) Directly from the eigenvalues. The inverse of the eigenvalues represent the timescales. These can be determined from the characteristic equation. \\
(3) From the statistical distribution of an ensemble.
We consider an ensemble of model clouds, each evolving along its own trajectory, and compute the resulting density of points in phase space. The equilibrium region is expected to have a higher density of trajectories if clouds spend a significant fraction of their evolution nearby.

\subsection{Phase-space speed}

For the dynamical system written in the non-dimensional variables $(r,v,w)$ (equations \ref{nd_energy_dissipation} -- \ref{nd_force_balance}), the phase-space speed is
\begin{equation}\label{phase_speed}
v_{\rm ph}
=
\left[
\left(\frac{dr}{d\tau}\right)^2
+
\left(\frac{dv}{d\tau}\right)^2
+
\left(\frac{dw}{d\tau}\right)^2
\right]^{1/2}.
\end{equation}

The first term is the radial velocity that is the accumulated structural
 response to earlier force imbalance.  The second and third terms are each expressible as the imbalance of force and energy, respectively, or geometrically
 as distances from their respective lines of equilibrium (nullclines).

We can show that near equilibrium, $dv/dt$ is proportional to the virial imbalance.
In equation \ref{nd_force_balance} for the restoring structural acceleration, 
the second term on the right-hand side can be written as the virial imbalance, so
\begin{equation}
\frac { dv } { d\tau } = -\frac{ v^2 } { r } + \frac { 3 } { \beta r_\ast} \Delta_{\rm vir} .
\end{equation}
Near equilibrium where $v$ is small and $r_\ast \approx 1$ the restoring force is directly proportional
to the virial imbalance.

Similarly, in equation \ref{nd_energy_dissipation} for $dw/d\tau$, the right-hand side is the energy imbalance.
so that $dw/d\tau = \Delta _{\rm energy}$. Then the phase speed is

\begin{equation}
v_{\rm ph}^2 = v^2 + \left( \frac { 3 } { \beta r_\ast} \Delta_{\rm vir} \right)^2 + \left( \Delta_{\rm energy} \right)^2  .
\end{equation}
Near the normalized equilirbrium, $r_\ast \approx 1$, $w_\ast \approx 1$, and $v \approx 0$. Therefore,
\begin{equation}
v_{\rm ph}^2 =\left( \frac { 3 } { \beta} \Delta_{\rm vir} \right)^2 + \left( \Delta_{\rm energy} \right)^2  .
\end{equation}
The phase speed is therefore proportional to the combined departures from force and energy balance.

The coefficient  $3/\beta$ in the virial-imbalance term reflects the mechanical stiffness of the self-gravitating cloud. For 
$\beta\sim1/3$ the coefficient is of order ten and larger than the coefficient of one on the energy imbalance. This indicates that a normalized departure from force balance produces a much stronger structural response than an equal normalized departure from energy balance, consistent with the more rapid establishment of force balance than the evolution in energy.

The trajectories in figure \ref{GE_KE} are color-coded by the phase-space speed. 
Near the equilibrium, where the trajectories slow as \hbox{$v_{\rm ph} \rightarrow 0$},
clouds spend a disproportionate amount of time even though the equilibrium is formally unstable.
The longer residence time near equilibrium explains why molecular clouds are commonly 
observed in states close to virial or hydrostatic balance despite their ultimate instability.

\subsection{Lifetime of the Near-Equilibrium State}

The development and persistence of approximate or near
equilibrium structure requires the timescale for relaxation to force balance be less
than the timescale for the growth of the energy instability, $t_{\rm force} < t_{\rm energy}$.
We can determine these characteristic timescales from the stable and unstable 
solutions of the characteristic equation \ref{E20}. The exact solutions are given as equations \ref{E27} and \ref{E28}.

The timescale $t_{\rm energy}$ is equal to the turbulent dissipation timescale, defined as a fraction of the crossing time by the
factor $\gamma \sim t_D/t_X$.
We can derive a limit on $\gamma$ that satisfies the condition on the timescales.
Write the roots of the characteristic equation as 
\begin{equation}
\lambda_1 > 0, \quad \lambda_{2,3} = -\alpha_s \pm i \varpi
\end{equation}
The real root $\lambda_1$ is the growth rate of the unstable mode. 
Therefore,
\begin{equation}
t_\mathrm{energy} = \frac { t_{x,0} } { \lambda_1}
\end{equation}
A perturbation in the stable structural mode behaves like,
\begin{equation}
\delta(\tau) \propto e^{ -\alpha_s\tau } \cos ( \varpi \tau + \phi)
\end{equation}
In the real part of the complex conjugate pair, $ -\alpha_s$ is the damping rate of the oscillation
while in the imaginary part, $\varpi$, is the rate of the oscillation itself.
This is the linearized version of the gravitational frequency of 
the cloud.
Therefore, the structural or force-response time is
\begin{equation}
t_\mathrm{force} = \frac { t_{x,0} } { \varpi }.
\end{equation}

At the normalized equilibrium $r_\ast = w_\ast = 1$, with the Larson-type driving $\alpha = 1/2$,
and the Lane-Emden structural value $\beta=1/3$, 
the coefficients of the characteristic equation (equations \ref{E21}) are
\begin{equation}
A = \frac{ 1 } { 2\gamma }, \quad B= \frac { 3 } {\beta }, \quad
C = -\frac { 3(1+2\alpha) } { 2\beta\gamma } = -\frac {9 } { \gamma } = -18A
\end{equation}

The condition on the time scales requires that $\varpi > \lambda_1$.
From Vieta's relations $\lambda_1 + \lambda_2 + \lambda_3 = -A $ 
and $\lambda_1\lambda_2 + \lambda_1\lambda_3 + \lambda_2\lambda_3 = B $
we have
\begin{equation}
\alpha_s = \frac { A + \lambda_1 } {  2 } 
\end{equation}
and
\begin{equation}
 \varpi^2 = 9 - \frac { A^2 } { 4 } + \frac { A\lambda_1 } { 2 } + \frac { 3 \lambda_2^2 } { 4 }
\end{equation}

Set $\varpi = \lambda_1$ as the limit of our condition. Eliminate $\lambda_1$,
\begin{equation}
2A^3 - 30A^2 + 135A - 270 = 0
\end{equation}
Since $A = 1/(2\gamma)$ we derive $\gamma_\mathrm{crit} = 0.05$.
Therefore the time scales in any physically relevant model with $\gamma$ of order unity
satisfy $t_\mathrm{force} < t_\mathrm{energy}$

Alternatively, if we adopt $\gamma=1$, then the unstable root
$\lambda_1 \approx 0.88$ and the oscillation frequency of the stable
pair $\varpi \approx 3.12$. Therefore,
\begin{equation}
\frac { t_{\rm energy} } { t_{\rm force} } \approx \frac { \varpi } { \lambda_1 } \approx 3.5
\end{equation}
This means that the rate to evolve to force balance is 3.5 times faster than the growth rate of the
instability in energy that ultimately destroys a cloud by collapse or expansion.

\subsection{Phase-Space Density of an Ensemble of Clouds}\label{phase_space_density}

Figure \ref{phase_density}  shows the distribution of an ensemble of model clouds in the $(\tilde{U},\tilde{K})$ plane, constructed by integrating a large number of trajectories initialized in the vicinity of the equilibrium point. 
For each trajectory, the cloud state is sampled at uniform intervals
in the non-dimensional time $\tau$ over the integration interval
$0<\tau<\tau_{\max}$. The phase-space density is then obtained by
binning all sampled points in the $(\tilde U,\tilde K)$ plane. Thus
the figure is  a residence-time histogram showing where trajectories spend a larger
fraction of their evolution. Equivalently, the distribution may be
interpreted as the cloud population that would be observed in a
single snapshot if the clouds were drawn with random ages from
the same ensemble of trajectories.

\subsubsection{Formation of overdensities along eigenvector directions}

Along the {stable eigenvector}, corresponding to the oscillatory modes, trajectories converge toward the equilibrium where their slower phase-space  speeds  lead to an accumulation and increase in the phase-space density along the stable manifold.
Along the {unstable eigenvector}, trajectories depart from equilibrium. However, the rate of motion varies along this unstable direction. Near equilibrium the phase-space speed is small, so clouds linger, resulting in a higher density near equilibrium. Farther away from equilibrium, the speed increases and the density decreases.

\subsubsection{Asymmetry of the unstable direction}  

\begin{figure*}[hbt!]
\includegraphics[width=6.5in,trim={0 0.0in 0.0in 0.0in},clip] {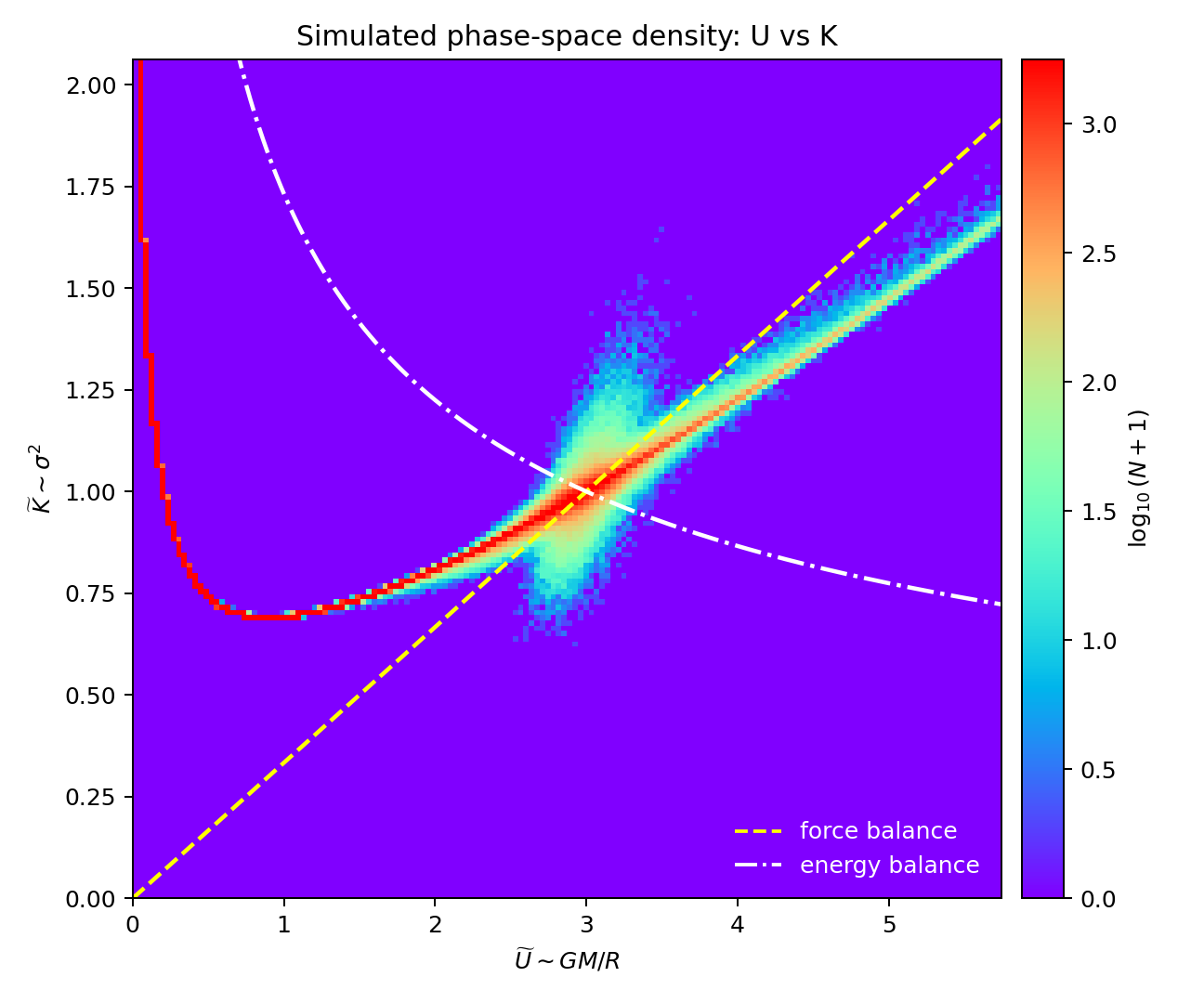}
\caption{Density of an ensemble of clouds in non-dimensional phase space as 
described in \S \ref{phase_space_density}. The trajectories 
are projected onto the plane of the kinetic versus gravitational energies.}
\label{phase_density}
\end{figure*}

A notable feature of figure \ref{phase_density}  is the asymmetry of the distribution along the unstable direction. The trajectories do not populate the collapse and expansion branches equally. Instead, there is a systematic bias toward the expansion side of the phase space.

This asymmetry arises from the scale dependence of the energy injection term. For $\alpha = 1/2$, the injection rate increases with scale in such a way that turbulent support grows more rapidly in expanding configurations than in contracting ones. 
In contrast, the exponent of 1/3 in the Kolmogorov cascade results in a constant injection rate across all scales.

As a result, expanding clouds receive relatively stronger energy input,
and collapsing clouds experience comparatively weaker support and evolve more rapidly.
Because the phase-space speed is higher in the collapsing branch, trajectories spend less time there. Conversely, the slower evolution along the expanding branch leads to a higher density. The net effect is an asymmetric population of the unstable manifold, favoring expansion over collapse. 
The trend is in the correct direction to explain the inefficiency of star formation. According to observational estimates, less than 1\% of
the mass of the molecular ISM is involved in collapse and star formation \citep{Evans_2021}.

\subsubsection{Observational selection effects}

The left region of the diagram corresponds to clouds with low gravitational energies that indicate large radii.  
This combination implies a low column or surface density.
These diffuse, extended structures would generally be classified observationally as low-level background emission rather
than clouds. 
Excluding this part of phase space improves the observable localization of clouds around equilibrium.

\subsubsection{Interpretation}

The phase-space density of an ensemble of clouds provides a statistical counterpart to the single-trajectory analysis. It shows that the near equilibrium state is not merely a transient feature of individual trajectories, but a state of high-probability when viewed over an ensemble.

This result reinforces the conclusion that the observed prevalence of molecular clouds in approximate equilibrium does not imply that the equilibrium is stable, only that more of the clouds spend more of their time  in its vicinity. The saddle-point structure, combined with scale-dependent energy injection and observational selection, naturally produces the observed clustering of equilibrium cloud properties.

\begin{figure*}[hbt!]
\includegraphics[width=6.5in,trim={0 0.0in 0.0in 0.0in},clip] {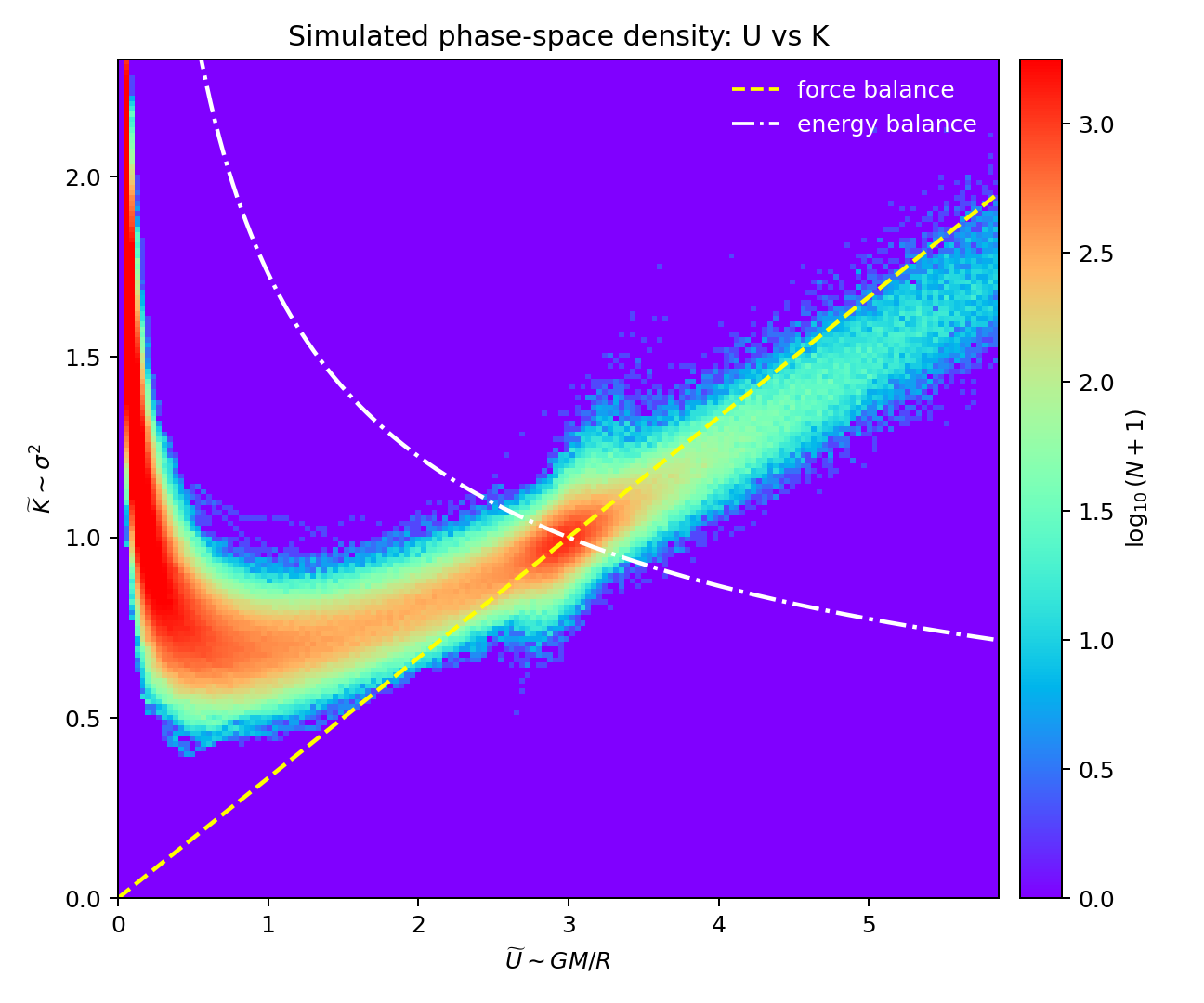}
\caption{Density of an ensemble of clouds in non-dimensional phase space as 
described in \S \ref{phase_space_density_noise}. The deterministic trajectories are the same as in figure 
\ref{phase_density} but spread by fluctuations in energy and pressure. }
\label{phase_density_noise}
\end{figure*}

\subsubsection{Phase-Space Density with Environmental Fluctuations}\label{phase_space_density_noise}

Molecular clouds do not evolve in isolation but within a turbulent field that introduces fluctuations on a range of spatial and temporal scales. 
We can represent these effects by adding stochastic perturbations implemented as Ornstein-Uhlenbeck (OU) processes  which provide a simple description of correlated noise with a finite coherence time \citep{Uhlenbeck_1930}. The OU formulation ensures that fluctuations evolve smoothly in time with a characteristic correlation timescale, in this case comparable to that of the surrounding turbulence.
Because these fluctuations act continuously as the cloud evolves, they perturb the trajectories in phase space, causing deviations from the deterministic paths described in the previous section.

Figure \ref{phase_density_noise} shows the phase-space density of an ensemble of model clouds in the $(\tilde{U},\tilde{K})$ plane whose evolution is subject to stochastic fluctuations in both external pressure and the supply of turbulent energy.  

The simulation shows that with stochastic perturbations, the overdensity near equilibrium is enhanced by diffusive confinement described as follows.
The random displacements of a trajectory increases its path length as the cloud evolves toward and through the region of near equilibrium
before becoming unstable. In this random walk, the time required
increases with the increased distance squared. Thus the turbulent fluctuations increase the residence time near equilibrium.

The simulation shows that environmental fluctuations can increase rather than diminish the phase-space density around equilibrium,
 as long as the fluctuations are not so severe that they disrupt the cloud.

 \section{Phase-Space Speed and the Condition for Hydrostatic Equilibrium}\label{phase_space_speed}

The condition for hydrostatic equilibrium is derived from the
Euler equation for an inviscid fluid with self-gravity,
\begin{equation}
\rho \frac { D {\bf v}  }   { Dt  } = -{\bf \nabla}P - \rho {\bf \nabla}\Phi
\end{equation}
by the requirement that the material derivative $D{\bf v}/Dt  =\partial_t {\bf v} + ({\bf v \cdot \nabla}) {\bf v} \rightarrow 0$. 
This requires that the bulk flow velocity $V$ be small compared to both the effective sound speed $a$ 
and the characteristic dynamical speed \citep{Keto_2025},
\begin{equation}\label{HEv}
V \ll a,
\qquad
V \ll \sqrt{gL},
\end{equation}

The second condition is nearly identical to the first. At equilibrium, the virial-force balance is
\begin{equation}
3\sigma_0^2 \approx \Gamma \frac {GM} { R }
\end{equation} 
where $g \sim \Gamma GM/R^2$ and $\Gamma$ is a factor of order unity
related to the density distribution. 
Therefore, the second requirement is $ V \ll \sqrt{gL} \sim \sqrt{3}\sigma_0 $.

The relevant quantity for comparison with the hydrostatic condition is the radial velocity variable
\begin{equation}
v = \frac{dr}{d\tau} = \frac{\dot{R}}{\sigma_0}.
\end{equation}

\begin{figure*}[hbt!]
\includegraphics[width=6.5in,trim={0 0.0in 0.0in 0.0in},clip] {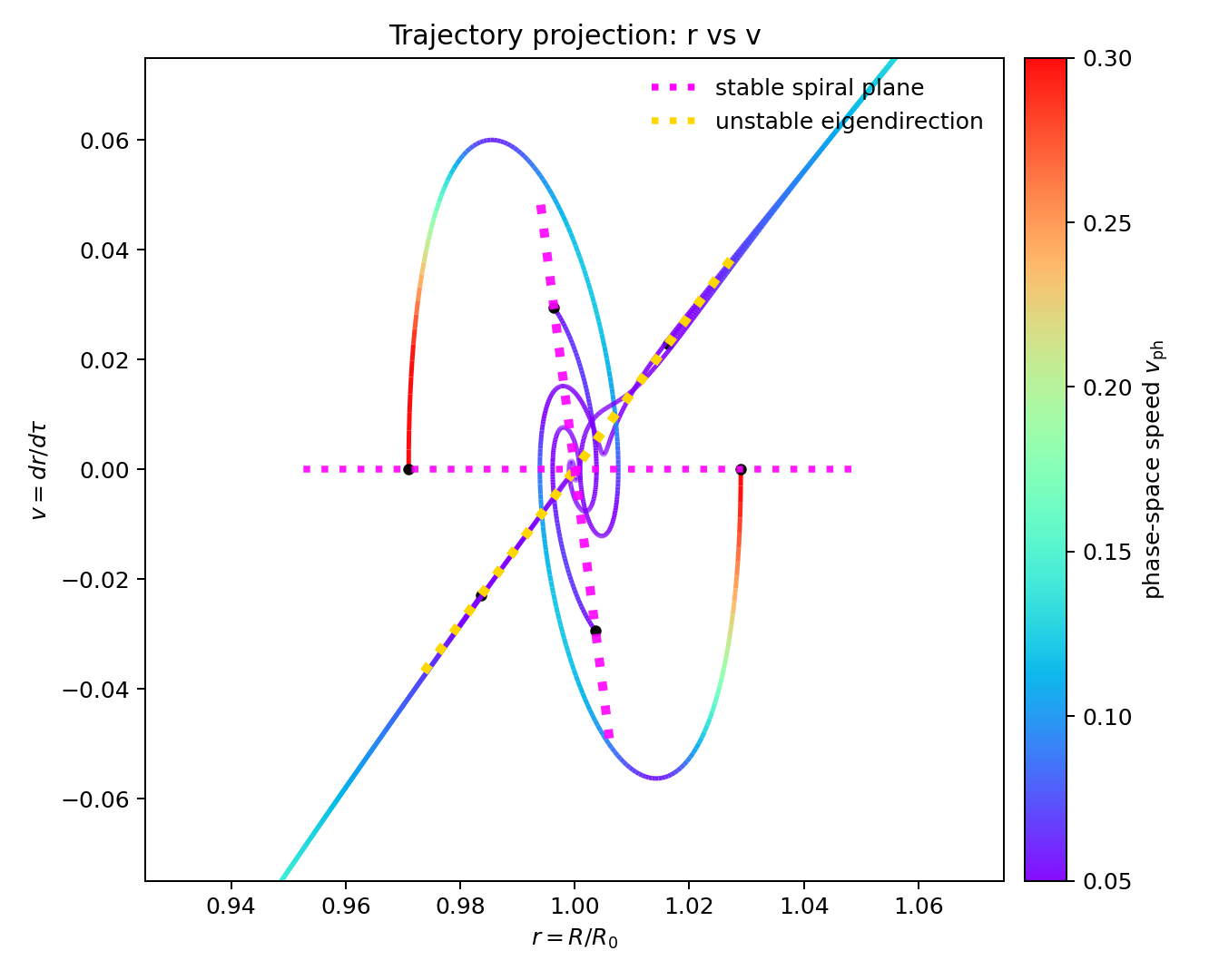}
\caption{Trajectories in non-dimensional phase space as described in \S \ref{phase_space_speed}. The trajectories 
are projected onto the plane of velocity versus radius.}
\label{r_v}
\end{figure*}

Figure \ref{r_v} shows the same calculation as in figure \ref{GE_KE} but now projected on the $(r,v)$ plane instead of $(\tilde{U},\tilde{K})$.
The figure provides a direct illustration of the condition required for hydrostatic equilibrium. 
The equilibrium point is located at $(r,v) = (1,0)$. 
In the vicinity of $r = 1$, all trajectories converge toward values
\begin{equation}\label{E71}
|v| \ll 1.
\end{equation}
This equation  is the dimensionless statement of equations \ref{HEv}.

Although $v$ describes the motion of the cloud boundary, the internal velocity field associated with homologous evolution scales as
\begin{equation}
v(r_{\rm int},t) \sim \frac{\dot{R}}{R}\, r_{\rm int},
\end{equation}
where $r_{\rm int}$ is any radius $< R$.
So the largest bulk velocities within the cloud are also of order $|\dot{R}|$.

In summary, the figure demonstrates that before the cloud evolves toward the instability in energy,
the trajectories pass through the region near equilibrium where $v$ is small and 
 the conditions required for hydrostatic equilibrium $V \ll a$ and $V \ll \sqrt{gL}$ are satisfied.
 
 This complements the result from the preceding sections. The hydrostatic condition describes
 the bulk velocities in the cloud. The smaller phase speed, $v_{\rm ph}$, near 
 equilibrium means that the force and energy imbalances are small, and that
 the whole cloud state is evolving more slowly near equilibrium. 
 
\section{Phase Lag and the Anti-Correlation of Column Density and Velocity Dispersion}

An interesting observational result for molecular clouds is the apparent anti-correlation between column density $\Sigma$ and the velocity dispersion $\sigma$ seen in representations such as figure~3 of \citet{Keto_2024}. This behavior arises naturally from the time-dependent dynamics near the equilibrium saddle point, and in particular from the phase lag between structural and energetic variables along spiral trajectories in phase space.

As shown in sections \ref{trajectories} and \ref{phase_space_speed},
the trajectories in phase space take the form of spirals toward or away from the equilibrium point.
Changes in radius $r$ occur on the dynamical timescale, while changes in turbulent energy (through $w$) occur on the dissipation timescale.
As a result, the extrema of $r$ and $w$ are offset in time, and there is a phase lag between structural evolution and energetic evolution.

Because
\begin{equation}
\Sigma \propto r^{-2},
\qquad
\sigma \propto \frac{w}{r},
\end{equation}
the phase lag between $r$ and $w$ produces a systematic relationship between the observables $\Sigma$ and $\sigma$ along a trajectory.

Consider a typical oscillation:

\begin{enumerate}

\item {Compression phase:}  
As the cloud contracts, $R$ decreases and $\Sigma$ increases. At the same time, gravitational compression increases the turbulent energy, so $\sigma$ rises, but with a delay.

\item {Maximum compression:}  
The radius reaches a minimum before the turbulent energy reaches its maximum. Thus the peak in $\Sigma$ precedes the peak in $\sigma$.

\item {Expansion phase:}  
As the cloud expands, $R$ increases and $\Sigma$ decreases. The turbulent energy continues to evolve due to dissipation, again lagging behind the structural change.

\end{enumerate}

Because of this lag, the trajectory in the $(\Sigma,\sigma)$ plane does not follow a single curve, but instead traces a loop. When projected onto the observational plane, this loop appears as an anti-correlation:

\begin{equation}
\Sigma \uparrow \quad \Rightarrow \quad \sigma \downarrow .
\end{equation}

This interpretation differs from a model that assumes a static scaling relation between $\Sigma$ and $\sigma$. Instead, the observed relation emerges from the {time-dependent dynamics of clouds near equilibrium}, not from a universal equilibrium condition.

This anti-correlation is generally present only in uninteresting clouds observed in unbiased surveys. 
Most targeted observations of clouds are meant to study
interesting phenomena within the clouds such as star formation.
Effects such as star-formation feedback generally increase the velocity dispersion. Since most star formation originates in
higher column densities, these observationally interesting clouds may show
a positive correlation of column density and linewidth.

 \section{Gravoturbulent and Scaling Interpretations as Projections of the Two-Dimensional Stability Surface}

\subsection{Gravoturbulent fragmentation as a one-dimensional slice}

The gravoturbulent fragmentation paradigm considers the formation and collapse of dense structures within a turbulent medium as the outcome of compression by converging flows \citep{Elmegreen_1993}. In this picture, localized regions of enhanced density arise in shocked layers or at stagnation points in the turbulent velocity field. Their subsequent evolution is determined by a comparison of their mass or surface density with a Jeans critical value.

The analysis in \citet{Elmegreen_1993} is explicitly concerned with the stability of compressed layers and the conditions under which the layers fragment gravitationally. The key result is that once a region exceeds the critical density for gravitational instability, it will collapse on a dynamical timescale, while a region with a compressed density below this threshold will expand and disperse. In this binary description there is no intermediate equilibrium state.

In this analysis, the turbulent velocity field is either prescribed or externally determined, so that the evolution is effectively described by a single structural variable, such as the density or radius. Thus the system is restricted to a one-dimensional trajectory in the two-dimensional phase space of macroscopic variables. The gravoturbulent framework can therefore be understood as a projection of the two-dimensional space onto a one-dimensional slice where the turbulent energy is held fixed or varies according to an assumed scaling relation.

The one-dimensional slice does not necessarily intersect the point of equilibrium that may be located elsewhere in
the full two-dimensional space. Since the energetic degree of freedom is not independently evolved, 
the system cannot evolve toward the force-balance and energy-balance equilibrium. 
Restricted to this slice, the full saddle structure is then reduced to a neutral-stability locus.
A  non-equilibrium initial condition starts either on one side or the other of the 
stability threshold, and the cloud evolves monotonically
toward dispersal or collapse.  
Thus, the absence of near equilibrium states in the gravoturbulent picture is not a general property of 
self-gravitating turbulence, but rather a consequence of restricting the dynamics to a single degree of freedom.

The conclusion that molecular clouds must eventually collapse or disperse is shared by all three descriptions
of self-gravitating turbulence, dynamical, thermodynamic, and gravoturbulent.
However, the gravoturbulent interpretation has been taken to imply that equilibrium or near equilibrium structures cannot develop within turbulent molecular clouds
\citep{MacLow_2004,Klessen_2005}. This stronger conclusion does not follow from the instability itself and is contradicted by both observations and the more complete two-dimensional  thermodynamic or dynamical systems stability analyses.

\subsection{The projection of the virial manifold}

A complementary observational perspective is provided by \citet{Heyer_2009}, who re-examined Larson's scaling relationships using improved observations of Galactic molecular clouds. They showed that Larson's size--line width relation,
\begin{equation}
\sigma_v \propto R^{1/2},
\end{equation}
is incomplete, and that the normalization of this relation depends systematically on the cloud surface density. 
Specifically, they derived (their eq.\ 11)
\begin{equation}
v_{\circ,G} \equiv \frac{\sigma_v}{R^{1/2}} 
= \left(\frac{\pi G}{5}\right)^{1/2} \Sigma^{1/2}
\end{equation}
for velocity dispersion $\sigma_v$, radius $R$, and surface density $\Sigma$.

This relation follows directly from the virial equilibrium condition,
\begin{equation}
2K + U = 0,
\end{equation}
and therefore expresses a balance between kinetic and gravitational energies. The key observational result of \citet{Heyer_2009} is that $v_{\circ,G}$ is not constant, but varies as $\Sigma^{1/2}$ indicating that clouds occupy a family of states rather than a single universal scaling relation.

In the framework developed here, this result can be interpreted geometrically. 
The two-dimensional
$(R,K)$ phase space is related to the three-dimensional virial-energy space $(K,U,\mathcal{P})$ as a reduced representation  
obtained by using the structural relation $U(R) \sim -GM^2/R$
and the boundary-pressure relation $\mathcal{P}(R,K) \sim K/R^3$.
The full virial equilibrium, including external pressure, defines a two-dimensional surface in the space of the three energy terms, kinetic, gravitational, and pressure, $K,\ U,\ \mathcal{P}$
subject to the constraint
$2K + U + \mathcal{P} = 0$.

 \citet{Heyer_2009} effectively project this surface onto the plane defined by $(\Sigma,\, \sigma_v/R^{1/2})$. In this projection, the normalization factor $v_{\circ,G}$ serves as a coordinate that traces the location of a cloud along the underlying two-dimensional virial manifold. The dependence $v_{\circ,G} \propto \Sigma^{1/2}$ is better understood as the projection of the virial equilibrium surface into observational variables rather than a scaling law.
This interpretation is consistent with the energy-space representation of \citet{Keto_2024} (their figure 5), where clouds occupy an elongated distribution in the $(K, U)$ plane aligned with the virial relation. Both of these representations of the observations are projections of the same underlying manifold, expressed in different coordinate systems.
Both the observational projections of 
$(K,U,\mathcal{P})$ and the dynamical evolution in 
$(R,K)$ developed here describe the same underlying equilibrium manifold in different coordinates.

\section{What about magnetic fields?}

The effect of magnetic fields in the hydrodynamics of a molecular cloud depends on the configuration and strength
of the field. The former is largely unknown and the latter only roughly estimated from a few examples. Nonetheless,
recent observations that compare the radial density structure of clouds to the structure predicted by the
Lane-Emden equation provide some guidance. 
The Lane-Emden equation combines an equation-of-state with
spherical hydrostatic equilibrium. We can derive constraints on the magnetic fields from
both.

The average density structure of Galactic and extragalactic clouds in M31 \citep{Keto_2024,Keto_2025} 
are both
well modeled by the Lane-Emden equation with an effective isothermal sound speed,
\begin{equation}
P_{\rm eff} =  P_{\rm turb}   + P_{\rm B,iso} \simeq \rho c^2_{\rm eff} 
\end{equation} 
where we omit the thermal pressure, $P_{\rm th}$, which  is negligible within supersonic turbulence.  

The hydrostatic condition requires the anisotropic magnetic stress to be dynamically small,
\begin{equation}
\left | {\bf \nabla \cdot} \left [ -\frac{1}{4\pi} \left ( B_iB_j - \frac{1}{3} B^2 \delta_{ij} \right ) \right ] \right | 
\ll| \nabla P_{\rm eff} |
\end{equation}
 over the region fit by the spherical model.
 One way this can happen is if the field is tangled on sufficiently small scales to have an approximately
 isotropic directional average,
 \begin{equation}
 \langle B_iB_j \rangle_\ell \simeq \frac{1}{3} \langle B^2 \rangle_\ell \delta_{ij} .
 \end{equation}
 Here we assume that the averaging scale, $\ell$, is between the scales of the magnetic correlation length
 and the cloud scale, $\ell_B \ll \ell \ll R$. The first inequality provides many fluctuating field elements 
 in an averaging volume; the second preserves variation on the cloud scale. In practice, the averaging
 length scale may be set by the observational resolution and data analysis procedure. The 
 hydrostatic condition imposes a constraint
 on the mean field within the averaging length scale.

We can now relate the magnetic pressure to the turbulent pressure with the reasonable assumption that the
magnetic and turbulent energies are in equipartition, $e_K \simeq e_B$. The hydrostatic condition requires that the
mean bulk velocity is small compared to the effective sound speed (equation \ref{HEv}). 
We decompose the velocity field into mean and fluctuating components, 
$u=\bar{u} + u^\prime$. We can assume that $u^\prime \sim \sigma$ and
that $\bar{u} \approx 0$. 

The turbulent kinetic energy derives from the fluctuating component,  
\begin{equation}
e_K = \frac{1}{2} \rho \langle u^{\prime 2} \rangle,
\end{equation}
and statistical isotropy results in
\begin{equation}
P_{\rm turb} = \frac{1}{3}  \rho \langle u^{\prime 2} \rangle = \frac{2}{3} e_K.
\end{equation}
If we define a mass-weighted mean velocity that may be
more appropriate for compressible turbulence, the relationship between
$P_{\rm turb}$ and $e_K$ is the same.

Assuming a similar decomposition of the turbulent magnetic field, $B = \bar{B} + b$, 
the fluctuating magnetic-energy density is,
\begin{equation}
e_B =  \frac{ \langle b^2 \rangle } { 8\pi } ,
\end{equation}
and the averaged Maxwell stress is,
\begin{equation}
\Pi^B_{ij} = \frac{ \langle b^2  \rangle } { 8\pi }\delta_{ij} - \frac{ \langle b_i b_j \rangle }{  4\pi} .
\end{equation}
The first term is the magnetic-pressure, and the second is the magnetic tension.
For statistically isotropic fluctuations,
\begin{equation}
\langle b_ib_j \rangle = \frac{1}{3} \langle b^2 \rangle \delta_{ij} ,
\end{equation}
and the tension term cancels two-thirds of the magnetic-pressure contribution, leaving
\begin{equation}
P_{B, \rm iso} = \frac{ \langle b^2 \rangle } {24\pi } = \frac13 e_B .
\end{equation}
Then
\begin{equation}
\frac { P_{B, \rm iso} } { P_{\rm turb} } = \frac { e_B/3  } { 2e_K/3  } = \frac12 .
\end{equation}

The combined isotropic support is consequently
\begin{equation}
P_{\rm eff}
=
P_{\rm turb}+P_{B,\mathrm{iso}}
\simeq
\frac{3}{2}P_{\rm turb},
\end{equation}
and
$c_{\rm eff}^2\simeq 3\sigma^2/2$. Under the additional
assumptions that the ordered field and anisotropic magnetic
surface stresses are dynamically small, the same support may be
represented approximately in the scalar virial theorem by
$K_{\rm eff}\simeq 3K_{\rm turb}/2$.

This scenario is consistent with but not uniquely required by the observed agreement of cloud structure with the 
Lane-Emden equation. This exercise demonstrates one way that a model that does not
explicitly include magnetic fields may be consistent with observed magnetized clouds.

\section{Summary and Conclusions}
This study addresses the apparent contradiction between the unstable, transient nature of molecular clouds and recent robust 
observations of hydrostatic structure within the clouds. 
Previous thermodynamic and dynamical systems analyses show that the equilibrium of self-gravitating molecular clouds 
in a two-dimensional phase space of structure (density or radius) and energy is a saddle point, stable in the
structural dimensional and unstable in the energy dimension. The contradiction is 
resolvable if the timescale to evolve
to force balance (hydrostatic equilibrium) is shorter than the timescale (inverse growth rate)
of the energy instability.

We study the evolution of molecular clouds as dynamical systems 
described by the time-dependent virial theorem and the
first law of thermodynamics along with a conservation equation for the energy lost
by turbulent dissipation and the energy transferred through the turbulent cascade.
From the resulting system of ordinary differential equations (ODEs), the Jacobian matrix confirms
the saddle structure of the equilibrium. More specifically, the structure is a saddle focus
or spiral saddle with the two stable modes oscillating on the gravitational frequency and
an unstable mode with a growth rate equal to the turbulent dissipation rate.
If the turbulent dissipation time equals the crossing time,
the eigenvalues indicate that the relaxation rate for the structural response, given by
the oscillation frequency is 3.5 time faster than the growth rate of the energy instability

The system of ODEs describes the evolution of a cloud through phase space. We find
that the trajectories  follow a typical pattern in the vicinity of a saddle point. The trajectories
first flow toward the equilibrium before departing in the unstable direction. The phase-space
speed along a trajectory is proportional to the combined magnitudes of the virial (force)
imbalance and the energy imbalance. By the conservation of flux in phase space,
the slowing as the cloud approaches equilibrium results in an overdensity in phase space
near equilibrium.

The combination of a relatively rapid relaxation rate toward structural (hydrostatic) equilibrium
and a predominance of observable clouds near equilibrium resolves the tension between the 
theoretical expectation of instability and the implied stability of hydrostatic and
virial equilibrium observed in molecular clouds.

The results of this study are consistent with the proposition that the observed equilibrium 
structure of molecular clouds is evidence
for the theoretically predicted structural stability of the metastable equilibrium. 
The observationally estimated lifetime of molecular clouds, approximately their
crossing time \citep{Elmegreen_2000}, is evidence for the theoretically predicted instability in the 
equilibrium energy balance with a comparable timescale.

 \bibliography{GRS_2024_08_27_REV}

@ARTICLE{Heyer_2009,
       author = {{Heyer}, Mark and {Krawczyk}, Coleman and {Duval}, Julia and {Jackson}, James M.},
        title = "{Re-Examining Larson's Scaling Relationships in Galactic Molecular Clouds}",
      journal = {\apj},
         year = 2009,
       volume = {699},
        pages = {1092-1103},
archivePrefix = {arXiv},
       eprint = {0809.1397},
 primaryClass = {astro-ph},
       adsurl = {https://ui.adsabs.harvard.edu/abs/2009ApJ...699.1092H}
}

@ARTICLE{Evans_2021,
       author = {{Evans}, Neal J., II and {Heyer}, Mark and {Miville-Desch{\^e}nes}, Marc-Antoine and {Nguyen-Luong}, Quang and {Merello}, Manuel},
        title = "{Which Molecular Cloud Structures Are Bound?}",
      journal = {\apj},
         year = 2021,
       volume = {920},
          eid = {126},
        pages = {126},
archivePrefix = {arXiv},
       eprint = {2107.05750},
 primaryClass = {astro-ph.GA},
       adsurl = {https://ui.adsabs.harvard.edu/abs/2021ApJ...920..126E}
}

@ARTICLE{Klessen_2005,
       author = {{Klessen}, Ralf S. and {Ballesteros-Paredes}, Javier and {V{\'a}zquez-Semadeni}, Enrique and {Dur{\'a}n-Rojas}, Carolina},
        title = "{Quiescent and Coherent Cores from Gravoturbulent Fragmentation}",
      journal = {\apj},
         year = 2005,
       volume = {620},
       pages = {786-794},
archivePrefix = {arXiv},
       eprint = {astro-ph/0306055},
 primaryClass = {astro-ph},
       adsurl = {https://ui.adsabs.harvard.edu/abs/2005ApJ...620..786K}
}

@ARTICLE{Keto_2020,
       author = {{Keto}, Eric and {Field}, George B. and {Blackman}, Eric G.},
        title = "{A turbulent-entropic instability and the fragmentation of star-forming clouds}",
      journal = {\mnras},
         year = 2020,
       volume = {492},
        pages = {5870-5877},
archivePrefix = {arXiv},
       eprint = {2001.02678},
 primaryClass = {astro-ph.GA},
       adsurl = {https://ui.adsabs.harvard.edu/abs/2020MNRAS.492.5870K}
}

@ARTICLE{Elmegreen_2000,
       author = {{Elmegreen}, Bruce G.},
        title = "{Star Formation in a Crossing Time}",
      journal = {\apj},
         year = 2000,
       volume = {530},
        pages = {277-281},
archivePrefix = {arXiv},
       eprint = {astro-ph/9911172},
 primaryClass = {astro-ph},
       adsurl = {https://ui.adsabs.harvard.edu/abs/2000ApJ...530..277E}
}

@ARTICLE{Elmegreen_1993,
       author = {{Elmegreen}, B.~G.},
        title = "{Star Formation at Compressed Interfaces in Turbulent Self-gravitating Clouds}",
      journal = {\apjl},
         year = 1993,
       volume = {419},
        pages = {L29},
       adsurl = {https://ui.adsabs.harvard.edu/abs/1993ApJ...419L..29E}
}

@ARTICLE{Lada_2025,
       author = {{Lada}, Charles J. and {Forbrich}, Jan and {Krumholz}, Mark R. and {Keto}, Eric},
        title = "{The Role of Pressure in the Structure and Stability of GMCs in the Andromeda Galaxy}",
      journal = {\apj},
         year = 2025,
        month = jun,
       volume = {986},
       number = {1},
          eid = {12},
        pages = {12},
          doi = {10.3847/1538-4357/adcf9d},
archivePrefix = {arXiv},
       eprint = {2501.16447},
 primaryClass = {astro-ph.GA},
       adsurl = {https://ui.adsabs.harvard.edu/abs/2025ApJ...986...12L}
}

@ARTICLE{Keto_2024,
       author = {{Keto}, Eric},
        title = "{Scales of Stability and Turbulence in the Molecular ISM}",
      journal = {Astronomische Nachrichten},
         year = 2024,
        month = nov,
       volume = {345},
          eid = {e20240044},
        pages = {e20240044},
          doi = {10.1002/asna.20240044},
archivePrefix = {arXiv},
       eprint = {2404.10979},
 primaryClass = {astro-ph.GA},
       adsurl = {https://ui.adsabs.harvard.edu/abs/2024AN....34540044K}
}

@ARTICLE{MacLow_2004,
       author = {{Mac Low}, Mordecai-Mark and {Klessen}, Ralf S.},
        title = "{Control of star formation by supersonic turbulence}",
      journal = {Reviews of Modern Physics},
         year = 2004,
        month = jan,
       volume = {76},
       number = {1},
        pages = {125-194},
          doi = {10.1103/RevModPhys.76.125},
archivePrefix = {arXiv},
       eprint = {astro-ph/0301093},
 primaryClass = {astro-ph},
       adsurl = {https://ui.adsabs.harvard.edu/abs/2004RvMP...76..125M}
}

@ARTICLE{Keto_2025,
       author = {{Keto}, Eric and {Lada}, Charles and {Forbrich}, Jan},
        title = "{Analytic Modeling of CO Surface-Density Profiles in the M31 Molecular Clouds}",
      journal = {\mnras},
         year = 2026,
        month = jun,
          doi = {10.1093/mnras/stag1169},
       adsurl = {https://ui.adsabs.harvard.edu/abs/2026MNRAS.tmp.1109K}
}

@ARTICLE{Uhlenbeck_1930,
       author = {{Uhlenbeck}, G.~E. and {Ornstein}, L.~S.},
        title = "{On the Theory of the Brownian Motion}",
      journal = {Physical Review},
         year = 1930,
        month = sep,
       volume = {36},
       number = {5},
        pages = {823-841},
          doi = {10.1103/PhysRev.36.823},
       adsurl = {https://ui.adsabs.harvard.edu/abs/1930PhRv...36..823U}
}

@ARTICLE{Klessen_2000,
       author = {{Klessen}, Ralf S.},
        title = "{One-Point Probability Distribution Functions of Supersonic Turbulent Flows in Self-gravitating Media}",
      journal = {\apj},
         year = 2000,
        month = jun,
       volume = {535},
       number = {2},
        pages = {869-886},
          doi = {10.1086/308854},
archivePrefix = {arXiv},
       eprint = {astro-ph/0001379},
 primaryClass = {astro-ph},
       adsurl = {https://ui.adsabs.harvard.edu/abs/2000ApJ...535..869K}
}

@ARTICLE{Padoan_2002,
       author = {{Padoan}, Paolo and {Nordlund}, {\r{A}}ke},
        title = "{The Stellar Initial Mass Function from Turbulent Fragmentation}",
      journal = {\apj},
         year = 2002,
        month = sep,
       volume = {576},
       number = {2},
        pages = {870-879},
          doi = {10.1086/341790},
archivePrefix = {arXiv},
       eprint = {astro-ph/0011465},
 primaryClass = {astro-ph},
       adsurl = {https://ui.adsabs.harvard.edu/abs/2002ApJ...576..870P}
}

@ARTICLE{Donkov_2025b,
       author = {{Donkov}, Sava D. and {Stefanov}, Ivan Zhivkov and {Kopchev}, Valentin},
        title = "{Thermodynamics of Fluid Elements in the Context of Turbulent Isothermal Self-Gravitating Molecular Clouds in Virial Equilibrium}",
      journal = {Universe},
         year = 2025,
        month = nov,
       volume = {11},
       number = {12},
          eid = {383},
        pages = {383},
          doi = {10.3390/universe11120383},
       adsurl = {https://ui.adsabs.harvard.edu/abs/2025Univ...11..383D}
}
 
 \end{document}